\documentclass[aps,pre, twocolumn, superscriptaddress,floatfix]{revtex4-1}
\usepackage{graphicx}
\usepackage{amssymb}
\usepackage{amsmath} 
\usepackage{float} 
\usepackage{subcaption} 
\usepackage[dvipsnames]{xcolor}
\usepackage{hyperref}
\usepackage{caption}
\usepackage{mathrsfs}

\begin{document}

\title {Self-organization and memory formation in two-dimensional jammed deformable matter under cyclic compression}

\author{Rahul Nayak}
\affiliation{The Institute of Mathematical Sciences, Taramani, Chennai 600113, India}
\affiliation{Homi Bhabha National Institute, Anushaktinagar, Mumbai 400094, India}
\author{Satyavani Vemparala}
\affiliation{The Institute of Mathematical Sciences, Taramani, Chennai 600113, India}
\affiliation{Homi Bhabha National Institute, Anushaktinagar, Mumbai 400094, India}
\author{Pinaki Chaudhuri}
\affiliation{The Institute of Mathematical Sciences, Taramani, Chennai 600113, India}
\affiliation{Homi Bhabha National Institute, Anushaktinagar, Mumbai 400094, India}


\begin{abstract}
We study the athermal mechanical response of deformable ring assemblies to quasistatic compression. Beyond jamming, further densification induces buckling of rings, resulting in macroscopic mechanical softening. Under cyclic compression, monodisperse systems anneal toward a nearly reversible path passing through an ordered state, whereas polydisperse systems converge to stable, hysteretic limit cycles. These limit cycles encode a robust memory of the training history that is retained even under subsequent overdriving. We show that macroscopic hysteresis in the disordered packings originates from directionally asymmetric non-affine deformations at the microscale while keeping contact network largely intact. Our findings demonstrate how particle deformability governs collective self-organization and memory formation in jammed soft matter.
\end{abstract}

\maketitle

\section{Introduction}

The jamming transition describes the emergence of rigidity in disordered collections of particles, such as grains, foams, and emulsions~\cite{liu1998jamming}. While the scaling laws and structural properties of jammed systems are well-established for particles modeled as point objects interacting via hard-core or soft repulsive potentials~\cite{OHern2003, LiuNagel2010, van2010jamming, Chaudhuri2010}, many physical and biological constituents, ranging from cells and microgels to polymeric rings, possess internal degrees of freedom that allow for significant shape deformations~\cite{pontani2012biomimetic, yoo2010polymer, manning2023essay}. Unlike rigid objects, these deformable entities adapt to local packing constraints according to their internal elasticity, altering the mechanical response of the assembly in ways that point-particle models cannot capture. The bulk behavior of these systems is therefore determined by both inter-particle contacts and the elastic energy stored in shape deformations, thereby linking particle-scale mechanics to the macroscopic response~\cite{Boromand2018, Treado2021}. The consequences of this coupling extend well beyond structural packing, viz. assemblies of deformable particles serve as model systems for studying glassy dynamics in highly compliant constituents~\cite{pellet2016glass, Gnan2019, gnan2021dynamical, ghosh2024onset, nayak2026glassy, bose2025understanding}, for modeling biological tissues~\cite{pasupalak2024epithelial, kumar2025particle}, and for understanding the rheology of squishy matter~\cite{poincloux2024rigidity, arora2024shape, ghosh2025two, bose2026study}. In regimes where particle deformations become large, the standard assumptions of granular mechanics, e.g. small strains, Hertzian contacts, and fixed particle shape, break down, requiring frameworks that explicitly couple particle-level shape change to the overall mechanical response~\cite{cantor2020compaction}. Crucially, in these deformable particle assemblies, individual particle identities and interstitial voids persist even at high densities, which requires modeling approaches distinct from confluent frameworks used to study dense cellular packings, where cell boundaries are shared and voids are entirely absent~\cite{bi2015density, henkes2020dense, sadhukhan2022origin}. This non-confluent distinction has consequences for localized stress transmission and contact network stability under driving. How the interplay between shape deformability and the persistence of interstitial voids influences the capacity of jammed assemblies to self-organize and encode memory under cyclic mechanical driving is our focus.

Disordered solids encode memories of their mechanical history, a phenomenon demonstrated across a range of systems, from sheared non-Brownian suspensions and bubble rafts to crumpled elastic sheets~\cite{Regev2013, Paulsen2014, mukherji2019strength, keim2019memory, shohat2022memory, adhikari2025encoding}. Atomistic and elastoplastic modeling have established that single and multiple memories can be encoded via cyclic driving, which acts as a mechanical annealing process that progressively drives the system toward stable reversible limit cycles~\cite{fiocco2014encoding, kumar2022mapping}. The encoded memory can, however, be fragile -- it is erased when the driving amplitude exceeds the training amplitude, and is ultimately bounded by the irreversibility transition, the critical amplitude beyond which the system enters a regime of plastic diffusion~\cite{mungan2025self}. Microscopically, memories are organized via localized plastic rearrangements around soft spots that, through training, become reliably activated and can be modeled as hysteretic two-state units; when interactions between these units become frustrated, they induce multi-periodic cycles and enable the emergence of complex memory behaviors~\cite{szulc2022cooperative, mungan2019networks}. The vast majority of these studies, however, concern systems where changes in particle shape play no role -- whether deformability introduces qualitatively new memory mechanisms remains an open question that this work addresses directly.

To explore these aspects, we model two-dimensional athermal assemblies of deformable soft matter as ring polymers and investigate their response to quasistatic compression and cyclic driving. Beyond jamming, individual rings buckle under compression, producing a compliant ``squishy jammed'' state marked by a pronounced softening of the bulk modulus. Under cyclic driving, these assemblies self-organize into stable limit cycles whose nature depends critically on structural disorder -- monodisperse systems undergo athermal annealing toward near-crystalline, nearly reversible paths, recovering memory through ordering, while polydisperse systems converge to persistent hysteresis loops encoding robust structural memory within a disordered state, stable against extreme overdriving and nested subcycles. This resilience originates in the topological stability of the self-organized contact network, preserved even under large post-buckling deformations. The macroscopic hysteresis arises from directionally asymmetric non-affine deformations that redistribute internal contact forces within a largely frozen network, a sub-topological irreversibility mechanism absent in rigid-particle systems. These findings reveal that the ability of rings to absorb compressive stress through shape change protects the self-organized topology under driving, giving rise to a class of highly compliant jammed assemblies capable of robustly encoding and reproducing trained mechanical states.

The remainder of this paper is organized as follows. In Sec.II, we  describe the ring polymer model and the three systems studied, as well as the quasistatic  compression protocol. Results are presented in Sec.III, followed by a discussion in  Sec.IV.

\section{Model and methods}
We consider a two-dimensional assembly of $N=1000$ deformable particles modeled as ring polymers~\cite{smrek2020active,ghosh2024onset}.
For the interaction between monomers, we used a combination of Weeks-Chandler-Andersen (WCA) potential
\begin{equation}
    U_{LJ}(r)=
    \begin{cases}
    4\epsilon[(\frac{\sigma_{m}}{r})^{12}-(\frac{\sigma_{m}}{r})^{6}]+\epsilon & \text{if }r\le 2^{\frac{1}{6}}\sigma_{m}\\
    0 & \text{if }r>2^{\frac{1}{6}}\sigma_{m}
    \end{cases}
\end{equation}
between all monomers, and finite extensible nonlinear elastic (FENE) potential
\begin{equation}
    U_{FENE}(r)=-\epsilon k_{F}R_{F}^{2}\ln [1-(\frac{r}{R_{F}\sigma_{m}})^{2}] \quad \text{if }r<R_{F}\sigma_{m}
\end{equation}
between bonded monomers, where $\sigma_{m}$ is the diameter of each monomer (and the unit of length), $\epsilon$ is the unit of energy, $k_{F}=15$ is the spring constant, and $R_{F}=1.5$ is the maximum extension of the bond ~\cite{smrek2020active}.
To model the bending flexibility of the ring backbone, we implement an angular potential acting on the angle $\phi$ between consecutive bond vectors:
\begin{equation}
    V(\phi)=K_{\phi}(1-\cos(\phi-\pi))
\end{equation}

We consider three model systems. (i) A monodisperse system (labelled $M0$) in which all rings have the same number of monomers ($n_m = 100$) and the same angular stiffness $K_\theta = 100$. (ii) A stiffness-polydisperse system (labelled $K30$) in which all rings have $n_m = 100$ monomers but differ in angular stiffness, with $K_\theta$ sampled from a uniform distribution of mean $\overline{K_\theta} = 100$. (iii) A size-polydisperse system (labelled $P30$) in which all rings have the same stiffness $K_\theta = 100$ but differ in the number of monomers, with $n_m$ sampled from a uniform distribution of mean $\overline{n_m} = 100$. In both polydisperse systems, polydispersity is defined as the ratio of the standard deviation of the distribution to its mean, expressed as a percentage; following our prior study~\cite{nayak2026glassy}, we use $30\%$ polydispersity in both cases.  Throughout this work, all energy, density, and pressure values are reported in reduced Lennard-Jones units.

Initial athermal configurations are generated by sampling equilibrium supercooled states at $T = 1.0$~\cite{nayak2026glassy}. These configurations are then instantaneously quenched to their respective inherent structures using conjugate gradient minimization. This sampling is performed at low densities where the resulting athermal states are unjammed and particles remain non-overlapping.

We employ a quasistatic protocol to investigate the mechanical response under compression. At each step, the simulation box length is rescaled by a factor $\alpha_1 = 0.9995$ for compression (or $\alpha_2 = 1/\alpha_1$, for decompression). Each rescaling is followed by an energy minimization using the conjugate gradient method until the system reaches a force-balanced state, defined by the norm of the global force vector falling below $10^{-7}$. This procedure ensures that the system remains in a local energy minimum throughout the compression-decompression cycles.

\begin{figure*}[t]
  \centering
    \includegraphics[scale=0.25]{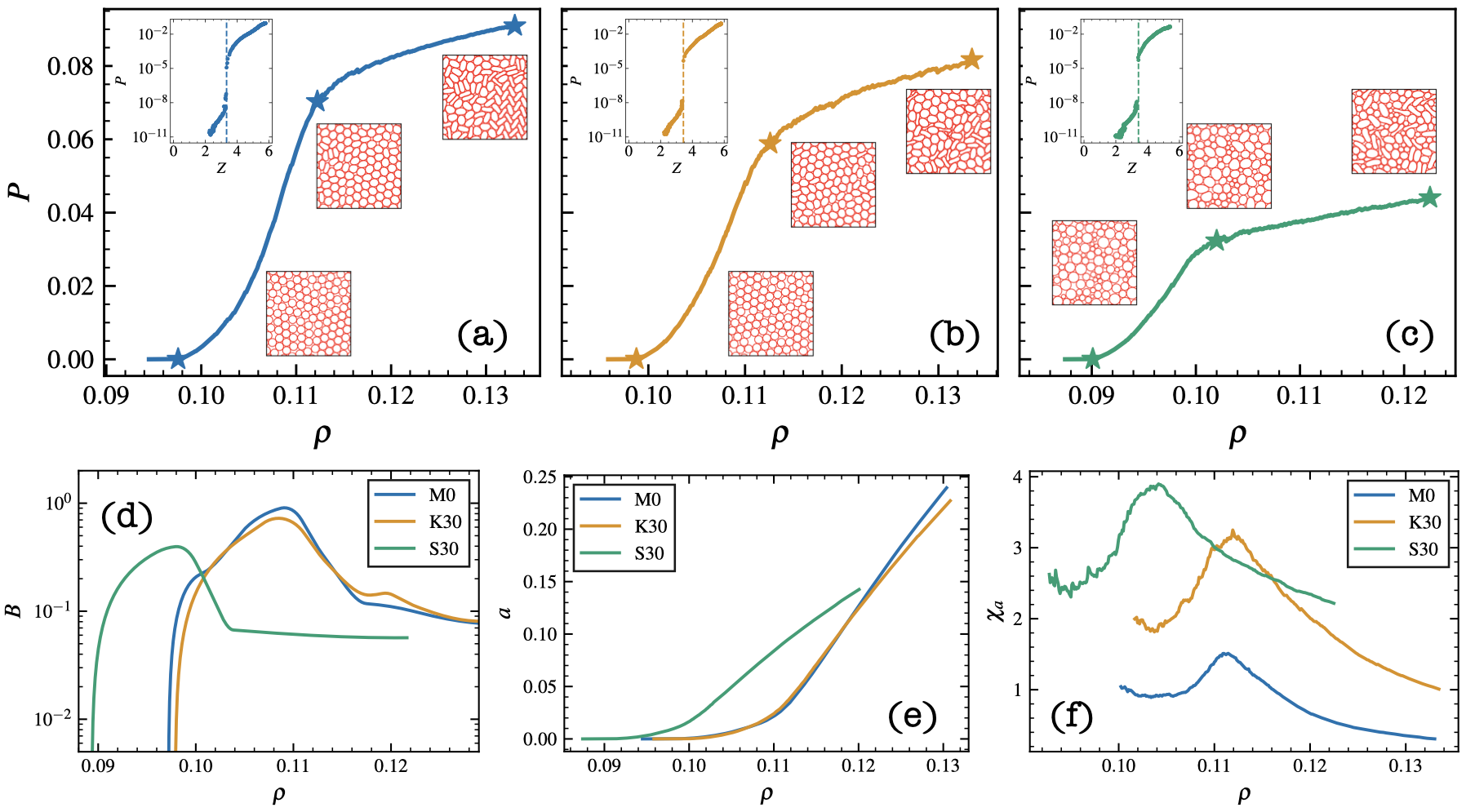}
    \caption{{\bf Response to compression.} Pressure ($P$) as a function of density ($\rho$) during athermal quasistatic compression of (a) monodisperse (M0), (b) stiffness-polydisperse (K30), and (c) size-polydisperse (S30) ring assemblies. Representative snapshots of mechanically stable configurations at the marked densities illustrate the progression of shape changes across the buckling regime. Insets in (a)--(c) show the evolution of pressure with mean contact number $Z$; the vertical line marks the estimated jamming threshold $Z_J$. (d)--(f) Corresponding measurements of bulk modulus $B$, mean asphericity $\langle a \rangle$, and asphericity susceptibility $\chi_a$, respectively, as functions of density $\rho$. The peak of $\chi_a$ identifies the buckling threshold $\rho_b$.}
  \label{fig1}
\end{figure*}

\begin{figure*}[t]
  \centering
    \includegraphics[scale=0.25]{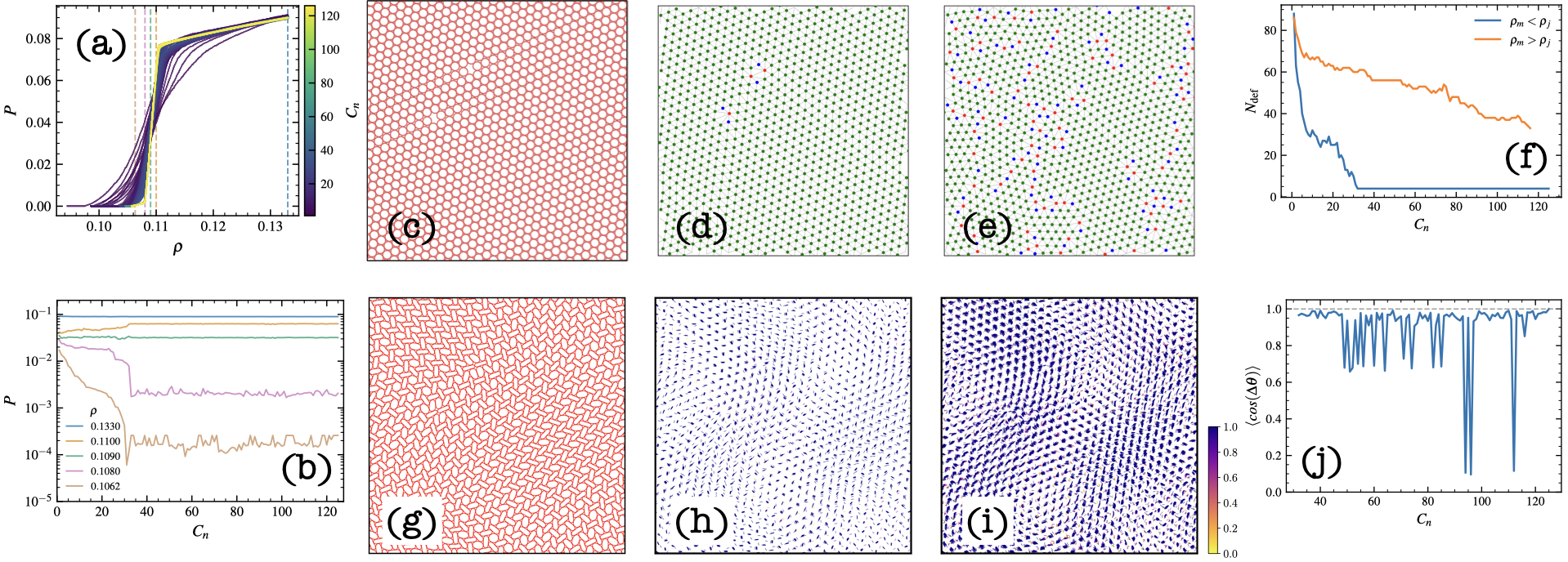}
    \caption{{\bf Response of monodisperse rings under cyclic compression.} (a) Pressure ($P$) as a function of density ($\rho$), showing compression and decompression branches over successive cycles; color indicates cycle number. (b) Pressure as a function of cycle number $C_n$ at selected densities indicated in (a). (c),(g) Representative configurations near asymptotic jamming ($\rho = 0.106$) and at the maximum density ($\rho_t = 0.133$) attained during cycling. (d),(e) Defect maps at $\rho = 0.106$ for the final and initial cycles, obtained via Voronoi analysis; red, green, and blue denote rings with 5, 6, and 7 neighbors, respectively. (f) Evolution of the number of defects, $N_{\mathrm{def}}$, at $\rho = 0.106$ with cycle number for two protocols: one with $\rho_m < \rho_J$ (as in (a)) and another with $\rho_m > \rho_J$. (h),(i) For the cycles shown in (a), superposition of configurations at $\rho = 0.133$ over the last 10 and 50 cycles, using a rod representation of the rings (see main text). (j) Evolution of the average change in ring orientation with cycle number, in the asymptotic limit.}
  \label{fig2}
\end{figure*}

\begin{figure*}[t]
  \centering
    \includegraphics[scale=0.275]{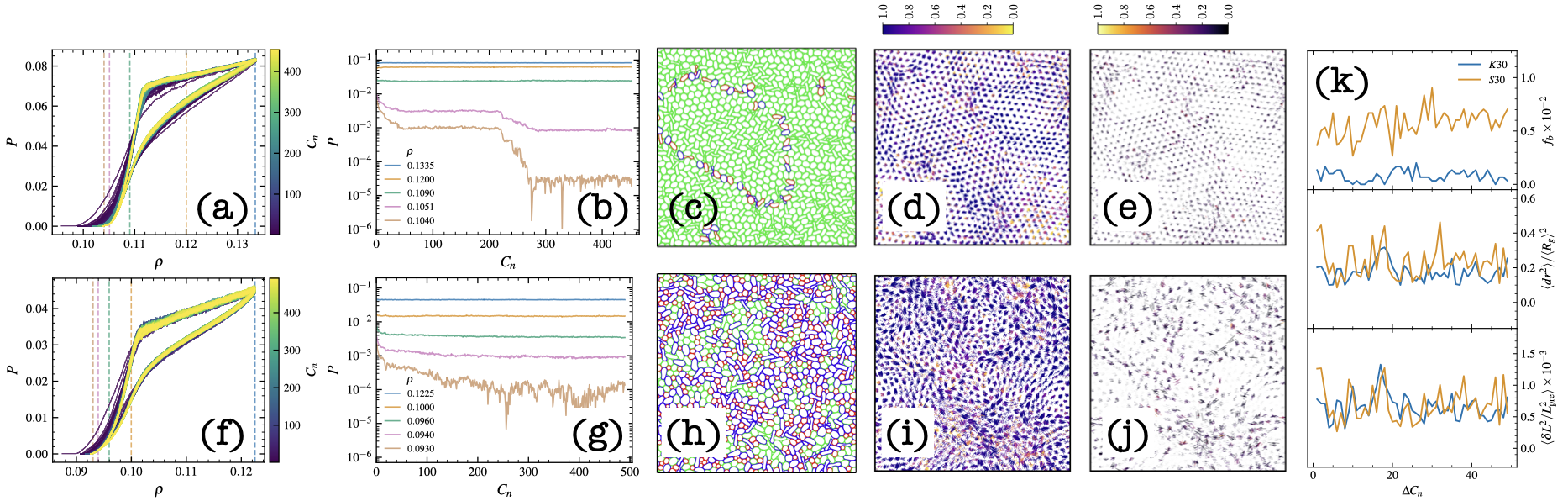}
    \caption{{\bf Cyclic compression response of polydisperse ring assemblies.} Top and  bottom panels correspond to stiffness- and size-polydisperse systems, respectively.
(a,f) Pressure $P$ as a function of density $\rho$ along compression and  decompression branches; color indicates cycle number $C_n$ (see colorbar).
(b,g) Pressure at selected densities marked in (a,f) as a function of $C_n$,  confirming convergence to an asymptotic limit cycle.
(c,h) Configurations in the asymptotic cycle at $\rho_t$, with rings colored by  coordination number $Z$: $Z < 6$ (red), $Z = 6$ (green), and $Z > 6$ (blue).
(d,i) Stroboscopic superposition of configurations at $\rho_t$ over the final 50  cycles, using a rod representation; color indicates the orientational deviation of  each ring from its orientation in the last frame, quantified by $\cos\Delta\theta$.
(e,j) Same as (d,i), with rod lengths scaled by the cycle-to-cycle change in asphericity $\Delta a$, reflecting shape fluctuations between consecutive frames. 
(k) Stroboscopic Voronoi network analysis at $\rho_t$: (top) fraction of broken Voronoi bonds $f_b$, (middle) mean squared displacement $\langle dr^2 \rangle$, and  (bottom) bond length fluctuations $\langle \delta L^2 \rangle$, all measured between  consecutive frames.}
  \label{fig3}
\end{figure*}

\begin{figure*}[t]
  \centering
    \includegraphics[scale=0.25]{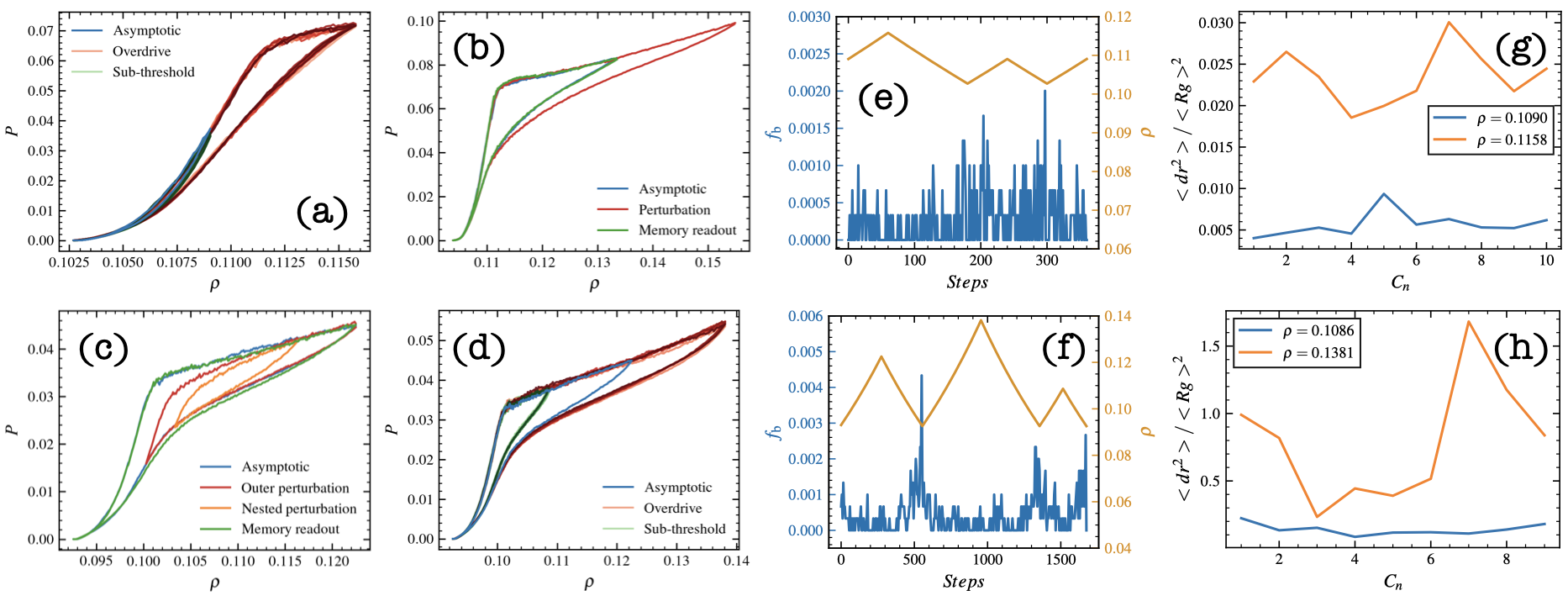}
    \caption{
    {\em {\bf Memory tests.} (Top)} Stiffness-polydisperse system.  (a) Alternation between pre-buckling ($\rho_t < \rho_b$) and post-buckling ($\rho_r > \rho_b$) cycles. 
    (b) Post-buckling hysteresis loops: the converged limit cycle trained at maximum density $\rho_t > \rho_b$ is overdriven by a larger-amplitude cycle ($\rho_r \gg \rho_b$) and subsequently re-evaluated at $\rho_t$. 
    {\em (Bottom)} Size-polydisperse system. 
    (c) Nested loops within the outer asymptotic limit cycle, explored with lower bounds greater than the initial unjamming density $\rho_m$. 
    (d) Sequence featuring a limit cycle at amplitude $\rho_{1m}$, followed by an overdriven cycle ($\rho_{2m} > \rho_{1m}$) and a subsequent subcycle of smaller amplitude ($\rho_{3m} < \rho_{1m}$). 
    (e, f) Fraction of Voronoi bonds broken, $f_b$, during the driving protocols in (a) and (d), respectively. These panels also display the instantaneous density accessed in each frame to correlate topological reorganization with specific points in the cycle. 
    (g, h) Stroboscopic mean squared displacement between consecutive frames, evaluated at the labeled densities, characterizing the positional stability of the states within the protocols shown in (a, d).
}
  \label{fig4}
\end{figure*}

\section{Results}

In an earlier study probing the thermal properties of these polydisperse ring systems~\cite{nayak2026glassy}, we established that increasing density drives slow dynamics through distinct mechanisms originating from different shape-change responses. In the stiffness-polydisperse system, all rings deform from circular to rod-like and form oriented bundles that drive the dynamical slowdown. On the other hand, in the size-polydisperse system, structural heterogeneity arising from a mixture of circular small rings and elongated large rings leads to frustration and glassy dynamics within an amorphous structure. Here, we examine the athermal mechanical response of the same systems under quasistatic compression and cyclic driving, where configurations are determined purely by mechanical stresses. We begin with the monodisperse system to assess the role of deformability, and then introduce stiffness and size polydispersity to examine how structural disorder modifies the mechanical response.

\subsection {Jamming and buckling}

We initiate compression from a unjammed configuration: the rings are circular (mean asphericity $\langle a \rangle = 0$; see Appendix for definition) and non-overlapping, and the pressure is negligible ($\approx 10^{-11}$). Under successive compression steps, the rings come into contact and the system eventually jams, marked by a rapid increase in pressure (Fig.~\ref{fig1}a--c) across all model systems studied. The jamming transition and the consequent onset of rigidity are also characterized by the bulk modulus ($B$) becoming finite (Fig.~\ref{fig1}d). Similarly, the mean contact number of the rings~\footnote{Two rings are deemed to be in contact if there is at least one monomer-monomer contact between them.}, $Z$, rises sharply at the transition (not shown), as expected at the onset of jamming. Notably, although the jamming density differs across the three systems, reflecting the dependence of the jamming threshold on the specific initial configurations used to initiate the compression protocol, the contact number at the onset of jamming $Z_J$ is approximately similar across systems: $3.334$, $3.435$, and $3.425$ for the monodisperse, stiffness-polydisperse, and size-polydisperse systems, respectively. As shown in the insets of Fig.~\ref{fig1}a--c, the pressure increases discontinuously with $Z$ at the estimated $Z_J$, confirming the onset of jamming. These values indicate that jamming in these systems is hypostatic, consistent with prior observations in frictionless freely-rotating granular chains~\cite{hoy2017jamming} and jammed packings of deformable particles~\cite{Treado2021}.

In all three cases, pressure increases monotonically with density after the onset of jamming. Beyond a threshold density $\rho_b$, however, the rate of increase of pressure with density decreases, leading to a non-monotonic behavior of the instantaneous bulk modulus $B = \rho\,(dP/d\rho)$ -- after the initial increase at jamming, $B$ drops sharply for $\rho > \rho_b$ (Fig.~\ref{fig1}d), signaling pronounced mechanical softening; in particular, the size-polydisperse system exhibits a notably smaller asymptotic modulus than the other two systems. The system nonetheless remains mechanically stable and transitions to a highly compliant, readily compressible (``squishy jammed'') state. This behavior is consistently observed across all three model systems, although $\rho_b$ varies between them. We identify the onset of this softening as ``buckling,'' which coincides with a sharp increase in the mean asphericity of the rings, $\langle a \rangle$ (Fig.~\ref{fig1}e), reflecting shape changes driven by increasing compressive stresses and the associated deviation of ring shapes from their preferred circular configuration~\cite{Boromand2018, Treado2021}. We observe that at similar pressures, structural disorder promotes larger $\langle a \rangle$, as the heterogeneous stress distributions, arising from the disordered contact topology, cause the most stressed rings to buckle preferentially, shifting $\rho_b$ to lower values relative to the ordered system. To locate the buckling threshold, we define a susceptibility $\chi_a = [\langle a^2 \rangle/\langle a \rangle^2 - 1]$ based on the normalized variance of $a$, which exhibits a pronounced non-monotonicity; its peak identifies $\rho_b$. Such buckling does not arise in well-studied point-particle models of jammed materials with purely soft repulsive interactions~\cite{OHern2003, zhao2011new}, nor in deformable particles with an incompressible core~\cite{cantor2020compaction}, highlighting that the observed softening requires both particle deformability and an unconstrained annular geometry that permits inward buckling under compression. Snapshots of the ring assemblies at selected densities (Fig.~\ref{fig1}a--c) illustrate the onset and progression of shape changes across the buckling regime. Further, a consequence of this shape-change due to compression is that there is a net decrease in ring areas, the packing fraction computed from instantaneous ring areas varies non-monotonically with density across $\rho_b$, rendering it an unreliable control parameter. We therefore use number density $\rho$ as the compression axis throughout the following discussion.

\subsection{Response to compression cycles}

Having established that the system becomes highly compliant and deformable beyond $\rho_b$, we examine whether cyclic compression induces structural reorganization leading to a limit cycle, and how this behavior depends on disorder by contrasting monodisperse and polydisperse systems. For each model system, we start from an unjammed state, compress to densities well above $\rho_b$, and then quasistatically decompress until unjamming. This cycle, characterized by the minimum ($\rho_m$) and maximum ($\rho_t$) accessed densities, is repeated multiple times, and convergence is assessed through macroscopic observables such as pressure and energy.

\subsubsection{Monodisperse rings}

Following the initial compression, quasistatic decompression eventually unjams the system with pressure becoming vanishingly small and $Z$ approaches zero albeit at a higher density than the jamming threshold, consistent with the well-known hysteresis at the jamming–unjamming transition~\cite{Chaudhuri2010}. Consequently, a full compression--decompression cycle produces a hysteresis loop, as observed in granular systems~\cite{bandi2013fragility}. Over successive cycles, we monitor $P$ as a function of $\rho$ (Fig.~\ref{fig2}a) and observe that with increasing cycle number ($C_n$), the area of the hysteresis loop progressively decreases~\cite{bandi2018training}, eventually approaching a limiting path (highlighted in yellow in Fig.~\ref{fig2}a) characterized by sharp changes in pressure at well-defined densities. The convergence to this asymptotic limit cycle is confirmed by monitoring the pressure at four representative densities along the compression and decompression branches (Fig.~\ref{fig2}b); for $C_n > 40$, the pressure at each density reaches a stable plateau. Furthermore, the jamming density $\rho_J$ increases progressively with each cycle, similar to behavior reported for frictional grains~\cite{bandi2013fragility}, and converges to a limiting value of $\rho_J \approx 0.1059$. Concurrently, the buckling density $\rho_b$ evolves toward an asymptotic value of $\approx 0.1091$. Thus, cyclic training results in a marked reduction of the separation between $\rho_J$ and $\rho_b$: from $\approx 0.0140$ in the initial cycle to $\approx 0.0032$ in the asymptotic state, signaling progressive self-organization of the mechanical response under periodic driving.

A striking feature of the asymptotic path is the emergence of an ordered triangular lattice formed by the rings at the point of jamming ($\rho_J = 0.1059$; Fig.~\ref{fig2}c), in contrast to the disordered initial configurations typical of thermalized supercooled states (Fig.~\ref{fig1}a). These initial configurations are characterized by a high defect density (Fig.~\ref{fig2}e), defined as rings with non-six-fold coordination [see Appendix for details], whereas the asymptotic state (Fig.~\ref{fig2}d) is nearly crystalline, featuring only isolated defect pairs associated with residual void pockets. Fig.~\ref{fig2}f illustrates the evolution of the defect population toward this low-density asymptotic limit. Successive compression cycles function as an athermal annealing process, progressively eliminating defects to drive the system toward order. The persistence of residual defects prevents the compression-decompression path from reaching perfect reversibility. Furthermore, convergence toward a crystalline structure promotes collective buckling, effectively narrowing the gap between the jamming and buckling thresholds. Defect removal efficiency is governed by the minimum density reached per cycle; access to the unjammed regime is essential, as it facilitates the transient void reorganization required for defect annihilation. If the system remains jammed throughout the cycle ($\rho_m > \rho_J$), this reorganization is hampered, signficantly slowing down convergence toward the defect-free limit (Fig.~\ref{fig2}f). Notably, crossing the buckling threshold is not strictly necessary for self-organized ordering; sub-buckling cycles are sufficient to drive annealing, indicating that the refinement is primarily governed by the unjamming-jamming transition. Nevertheless, accessing the post-buckling, shape-changing regime significantly accelerates the annealing process.

To further characterize the asymptotic path, Fig.~\ref{fig2}g shows the configuration at the maximum density attained during compression. This state results from compressing the ordered structure shown in Fig.~\ref{fig2}c -- as compression proceeds, the hexagonal motif becomes distorted, with the surrounding six rings growing more aspherical than the central one, a configuration made possible by the ability of the rings to buckle and deform. The final tiling is not perfect; owing to incommensuration with the square simulation box and the presence of residual voids in the initial ordered state, the assembly develops orientational domains separated by defect lines. Upon decompression, the ordered structure is recovered in the vicinity of unjamming. A closer examination of configurations at maximum density reveals an additional feature: stroboscopic superposition over the last 10 cycles (Fig.~\ref{fig2}h) confirms a cycle-invariant positional response, with each domain internally frozen by its coherent contact network. Over longer windows (last 50 cycles, Fig.~\ref{fig2}i), the stroboscopic superposition blurs inhomogeneously, more pronounced at the domain boundaries where rings undergo coupled translations and rotations, and less so in the frozen domain interiors. We quantify the associated fluctuations via the system-averaged orientational change $\langle\cos{\Delta\theta}\rangle$ relative to the last accessed state (Fig.~\ref{fig2}j), which reveals intermittent collective rearrangements via soft-modes enabled by the geometric frustration at the interface between domains. The asymptotic state thus has two distinct levels of order -- nearly rigid ordered domains where fluctuations are small, and geometrically frustrated boundaries that remain comparatively soft, supporting long-wavelength coupled translations and rotations as slowly evolving degrees of freedom even in the asymptotic cycle. Crucially, these reorientations occur only when $\rho_m < \rho_J$; if the cycle remains within the jammed sector, we have observed that the fluctuations are negligible.

\subsubsection{Polydisperse rings}

Unlike the monodisperse case, where cyclic compression annealed the system toward a nearly reversible limit cycle, the polydisperse systems converge to a qualitatively different asymptotic state: a persistent, stable hysteresis loop in the pressure--density plane (Fig.~\ref{fig3}a,f, yellow curves). This convergence to a limit cycle is confirmed by monitoring the pressure evolution at fixed densities (Fig.~\ref{fig3}b,g), which saturates at large $C_n$, for both systems.  Voronoi analysis of the asymptotic state at maximum density (Fig.~\ref{fig3}c) reveals that the stiffness-polydisperse system develops a coarse-grained granular structure separated by defect lines. These grains emerge from polycrystalline order developing near jamming, with grain interiors populated by the same distorted hexagonal motifs observed in the monodisperse case; however, unlike the nearly defect-free ordered state of the monodisperse  system, stiffness heterogeneity affects motif formation and thus frustrates spatial organisation and arrests the tiling at the grain level. In contrast, at the largest accessed density, the size-polydisperse system remains inherently amorphous in the asymptotic state, with defects distributed homogeneously throughout the assembly (Fig.~\ref{fig3}h), as the diversity of ring sizes produces different effective stiffnesses, leading to heterogeneous shape responses that suppress coherent structuring.

As in the monodisperse case, the jamming density $\rho_J$ shifts toward higher values with increasing $C_n$, converging to asymptotic values of $\rho_J \approx 0.1040$ and $\rho_J \approx 0.0929$ for the stiffness- and size-polydisperse systems, respectively. The respective asymptotic buckling densities ($\rho_b \approx 0.1118$ and $\rho_b \approx 0.1012$), by contrast, remain largely invariant. In the monodisperse system, progressive ordering narrows the distribution of local buckling thresholds, sharpening the collective onset and shifting $\rho_b$; in the polydisperse systems, intrinsic structural heterogeneity maintains a broad distribution of local thresholds despite the cycling, leaving $\rho_b$ insensitive to the structural reorganization.

To assess the stability of the asymptotic states, we perform stroboscopic analyses at maximum density. Cloud plots constructed from rod representations of the rings (Fig.~\ref{fig3}d,i), accumulated over the last 50 cycles, show that centers of mass remain largely localized. This localization is further supported by Voronoi network analysis, which reveals that the fraction of Voronoi bonds breaking between consecutive frames ($f_b$) is very small (Fig.~\ref{fig3}k, top): approximately $0.1\%$ for the stiffness-polydisperse system and $0.6\%$ for the size-polydisperse system. The lower bond breakage in the stiffness-polydisperse system reflects the greater mechanical stability of its granular skeleton. In the size-polydisperse system, the absence of any preferred local structural motif leaves a broader distribution of contact stabilities, making the network more susceptible to local contact rearrangements; however, these events remain rare and spatially localized, leaving the global network topology and the structural memory encoded by cycling intact. Despite the robustness of the contact networks self-organized through cyclic compression, residual fluctuations are nonetheless present: the stroboscopic mean squared displacement of Voronoi nodes ($\langle dr^2 \rangle$) remains centered around $0.2\langle R_g^2 \rangle$ (Fig.~\ref{fig3}k, middle), and bond length fluctuations ($\delta L^2$) for topologically intact bonds remain below $0.001$ (Fig.~\ref{fig3}k, bottom). These small but persistent geometric fluctuations within the fixed contact network reflect a degree of structural floppiness, stemming from localized orientational rearrangements and slight cycle-to-cycle variations in particle shape (Fig.~\ref{fig3}e,j). As a result, the asymptotic hysteresis loop broadens slightly from cycle to cycle, suggesting that the system explores a narrow range of mechanically stable states, consistent with multistability within the frozen contact network.

\begin{figure*}[htbp]
  \centering
  \includegraphics[scale=0.35]{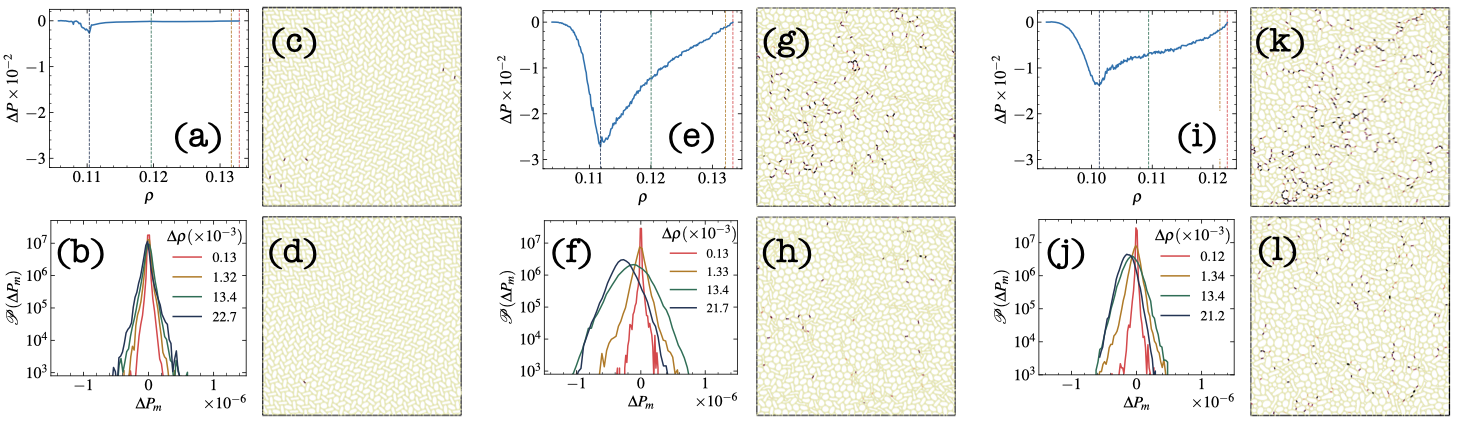}
   \caption{{\bf Hysteresis analysis.} 
   (a) Pressure difference $\Delta P$ between the decompression and compression branches of the asymptotic limit cycle for the monodisperse system, as a function of density $\rho$. (b) Distribution of the monomer-level pressure difference $\mathscr{P}(\Delta P_{m})$ sampled at selected density offsets $\Delta\rho$ from $\rho_t$, corresponding to the densities marked in (a). (c) Particle-scale non-affine displacement $D^2_{\min}$ during forward compression from $\rho_t - \Delta\rho$ to $\rho_t$. (d) Same as (c), measured during reverse decompression from $\rho_t$ to $\rho_t - \Delta\rho$. (e--h) Equivalent measurements to (a--d) for the stiffness-polydisperse system. (i--l) Equivalent measurements to (a--d) for the size-polydisperse system.
}
  \label{fig5}
\end{figure*} 

\subsection{Testing for memory formation}

The convergence to a stable limit cycle is indicative of memory formation and trainability. We therefore perform a series of standard tests to assess the robustness of this memory and to identify its physical origins (Fig.~\ref{fig4}). The top and bottom rows correspond to representative results for stiffness- and size-polydisperse systems, respectively.

In Fig.~\ref{fig4}a, the stiffness-polydisperse system is first trained at a density $\rho_t < \rho_b$, after which it is subjected to alternating cycles between $\rho_t$ and a higher density $\rho_r > \rho_b$. The system converges to a dual-stable response that reproduces both the trained cycle and the larger-amplitude cycle, indicating that positional memory is not only robust under alternating driving but remains undisturbed by the onset of buckling. Dynamic visualization clarifies the microscopic origin of this stability: the cores of the polycrystalline grains remain invariant throughout the process, despite the distortions induced by compression. We substantiate this topological robustness by measuring the fraction of Voronoi bonds broken between consecutive compression steps ($f_b$), shown in Fig.~\ref{fig4}e alongside the instantaneous density. The fraction is negligible throughout, with bond breakage occurring predominantly near the unjamming threshold. Although increased compression triggers localized fluctuations in bond-orientational order at the grain boundaries, the near-total absence of T1 topological rearrangements, with compressive stresses accommodated via ring shape changes rather than contact rewiring, ensures that the contact network of the trained structure is preserved. Consequently, the system can explore the post-buckled regime without losing the structural information encoded during training. 

We further probe memory stability for a system trained in the post-buckling regime, subjecting it to significant ``overdriving'' to determine whether high-amplitude deformations ($\rho_r \gg \rho_b$) can disrupt the encoded state, a phenomenon frequently reported in other amorphous solids where high-amplitude driving typically erases the memory of smaller training cycles~\cite{shohat2022memory}. As shown in Fig.~\ref{fig4}b, a system trained at $\rho_t > \rho_b$ is compressed to a significantly larger density, extending deep into the post-buckled regime, before being decompressed and cycled back to $\rho_t$. The structural memory is robustly retained despite transient deviations during which the decompression path from the overdriven state does not coincide with the training cycle; the system accurately recovers the training cycle on the way to unjamming and retraces the original compression path as it returns toward $\rho_t$.

The size-polydisperse system exhibits similar robustness. Figure~\ref{fig4}c demonstrates return-point memory: the system is trained at a maximal density $\rho_t$ and subsequently subjected to nested perturbations. Specifically, the decompression cycle is reversed at an intermediate density $\rho_m^{(2)} < \rho_b$, followed by an inner loop where compression is halted at a reduced maximum density ($\rho_t^{(2)} < \rho_t$) and then reversed at a higher minimum density ($\rho_m^{(3)} > \rho_m^{(2)}$). The system recovers its original response upon returning to the primary cycle, confirming strong memory retention. Figure~\ref{fig4}d examines the trained system under a distinct protocol consisting of alternating cycles of large-amplitude overdrive ($\rho_r^{(1)} > \rho_t$) and smaller-amplitude sub-cycles ($\rho_r^{(2)} < \rho_t < \rho_r^{(1)}$). The system converges to these stable cycles and ultimately retraces the trained path, demonstrating the global stability of the self-organized state. The fraction of Voronoi bonds broken during these perturbative cycles (Fig.~\ref{fig4}f) remains negligible, confirming the resilience of the topological network. We additionally compute the stroboscopic mean squared displacement $\langle dr^2 \rangle$ at the extreme densities reached during the perturbation cycles, viz., fluctuations remain minimal at the lower density threshold, whereas at the upper density limit, nodes undergo displacements comparable to $R_g$ without significant network reorganization, consistent with the structural floppiness within the frozen contact network identified in the asymptotic state analysis—a feature characteristic of the size-polydisperse system.

These results establish that asymptotic self-organized states in polydisperse systems are intrinsically stable under cyclic driving, despite the structural disorderedness, retaining robust memory across a wide range of perturbations. However, even though there is significant topological resilience, the systems continue to exhibit pronounced hysteresis at the macroscopic level, which we investigate next.

\subsection{Asymptotic hysteresis}

To quantify the path-dependence of the asymptotic limit cycle, we compute the pressure difference $\Delta P$ between the decompression and compression branches as a function of density (Fig.~\ref{fig5}a,e,i). In all cases, $\Delta P$ is negative throughout and reaches a minimum of approximately $-0.002$ in the vicinity of $\rho_b$ for the monodisperse system, with the polydisperse systems showing values an order of magnitude more negative, reflecting the fact that the monodisperse system converges to a nearly reversible cycle while the polydisperse systems retain a persistent loop with finite area. The pressure difference remains pronounced throughout the post-buckling regime in polydisperse systems, accounting for the persistent hysteresis in the pressure--density curves, whereas it is relatively negligible in the monodisperse case. A finite $\Delta P$ is also present in the pre-buckling regime, albeit over a smaller density window between $\rho_J$ and $\rho_b$.

Given the topological robustness of the asymptotic contact network discussed earlier, understanding the source of the pronounced hysteresis in the polydisperse systems requires a closer look. While we do observe some network rewiring, it is intermittent, involves only a small fraction of bonds per cycle, and does not produce the cascading rearrangements that typically drive hysteresis in rigid-particle amorphous solids. The irreversibility must therefore reside primarily at the sub-topological level, in the monomer-level configurations within the largely frozen contact network. To address this, we move beyond macroscopic observables and examine how stresses differ at the monomer scale between the decompression and compression branches. In Fig.~\ref{fig5}b,f,j, we plot the distribution of the local stress difference, $\mathscr{P}(\Delta P_{m})$, measured during the forward and backward paths, at densities marked in Fig.~\ref{fig5}a,e,i, with $\Delta\rho$ being the distance from $\rho_t$ and $P_m$ the hydrostatic stress measured at the monomer level. For monodisperse assemblies, the distributions are symmetric and centered near zero. In contrast, for polydisperse systems, the distributions become increasingly asymmetric and develop a finite mean for larger $\Delta\rho$, directly linking macroscopic hysteresis to a directional asymmetry in monomer-scale stresses and reflecting a redistribution of internal contact forces that depends on the driving direction, different from the nearly reversible response of the ordered monodisperse system.

We examine the particle-level origin of these stress asymmetries by computing the non-affine displacements, $D^2_{\text{min}}$, at a fixed distance $\Delta\rho = 1.3 \times 10^{-3}$ from the end-point $\rho_t$. In Figs.~\ref{fig5}c and d, we show these maps for the forward and reverse directions of the monodisperse system, respectively; while there are sparse locations of non-affine motion in the forward direction, these become negligible during decompression, confirming that the reverse path is nearly affine. Conversely, for the polydisperse systems (Figs.~\ref{fig5}g, h and k, l), non-affine motion is far more extensive in both directions, with a clear directional asymmetry. Significantly fewer non-affine events occur during decompression than compression, consistent with the system relaxing toward a local minimum on the decompression branch. The directional asymmetry of these non-affine events points to a rough local energy landscape in which compression and decompression traverse different paths. This localized irreversibility establishes the microscopic basis for the observed macroscopic hysteresis -- while the contact network defined by ring centers remains statistically robust across cycles, with only a small fraction of topological rearrangements per cycle, the monomer-level configuration is continually reshaped by these directionally asymmetric non-affine deformations, resulting in the loss observed in the pressure--density cycles. These findings indicate an energy landscape characterized by broad metabasins, within which the global topology of the contact network is robustly preserved, containing a finer structure of local minima associated with monomer-level reconfigurations. While the system remains stroboscopically confined to the topological basin of the contact network self-organized through cyclic driving, it explores the internal roughness of this landscape during each cycle, establishing these sub-topological dynamics as the source of observed asymptotic hysteresis in the learned state.

\section{Concluding summary}

Using extensive numerical simulations, we have investigated athermal mechanical response of deformable ring assemblies under quasistatic compression and cyclic driving across three compositions: monodisperse, stiffness-polydisperse, and size-polydisperse. Compression beyond the jamming threshold causes ring buckling, which decreases the bulk modulus as compressive stress shifts from inter-particle contacts to intra-particle shape deformations. Under cyclic driving, the assemblies evolve into steady states determined by their structural disorder. Monodisperse systems undergo athermal annealing via cyclic jamming--unjamming transitions, eliminating structural defects and converging to a reversible path through an ordered state. Polydisperse systems exhibit persistent hysteresis and settle into stable limit cycles, with asymptotic structures differing in spatial organistion, viz. the stiffness-polydisperse system forms a granular tiling with defect-line boundaries, whereas the size-polydisperse system remains amorphous; yet both exhibit robust memory retention across multiple levels of driving amplitude, including nested cycles and strong overdriving.

Thus, despite being a disordered assembly of discrete rings, the learned networks display a great degree of squeezability and recovery --- we link this robustness to strong topological stability, with only a small but finite fraction of contact rearrangements per cycle and no cascading plasticity. The absence of cascades reflects the role of deformability as a mechanical buffer --- compressive stress is absorbed through ring shape change before it can propagate through the contact network. Our analysis of local stresses and non-affine displacements establishes that hysteresis originates primarily from directionally asymmetric rearrangements within the largely frozen network, i.e., a sub-topological irreversibility mechanism encoded in ring-level shape deformations rather than contact rewiring. This picture invokes an energy landscape of broad metabasins defined by the contact network topology, each containing finer local minima associated with monomer-level reconfigurations.

The occurrence of robust structural memory along with high mechanical compliance identifies deformable particulate assemblies as a distinct class of trainable jammed materials, where stable topology and internal configurational fluctuations operate on separate levels of organization. The ability to absorb compressive stress through shape change protects the self-organized topology under driving, suggesting a design principle for soft materials that combine mechanical compliance with robust memory.

\section*{Acknowledgment}
We thank the HPC facility at The Institute of Mathematical Sciences for computing time. PC and SV also acknowledge support via the sub-project on the Modeling of Soft Materials within the IMSc Apex project, funded by Dept. of Atomic Energy, Government of India.

\section{Appendix}

\subsection{Asphericity and rod representation}

The shape changes of the rings are quantified by measuring the asphericity, calculated via the gyration tensor $S_{mn}$~\cite{ghosh2024onset, ghosh2025two}. For a ring of $N$ monomers, the gyration tensor is defined as:
\begin{equation}
S_{mn} = \frac{1}{N}\sum_{i=1}^{N}(r_{m}^{(i)}-r_{m}^{(CM)})(r_{n}^{(i)}-r_{n}^{(CM)}),
\end{equation}
where $m$ and $n$ denote Cartesian indices and $r_{m}^{(CM)}$ is the $m$-th component of the center-of-mass position. In two dimensions, this tensor yields two eigenvalues, $\lambda_{1}^2$ and $\lambda_{2}^2$ (where $\lambda_1^2 \geq \lambda_2^2$). We define the dimensionless asphericity $a$ as:
\begin{equation}
a = \frac{(\lambda_{1}^{2}-\lambda_{2}^{2})^2}{(\lambda_{1}^{2}+\lambda_{2}^{2})^2}.
\end{equation}
By this definition, $a=0$ for a perfect circle, with increasing values of $a$ indicating a progressive deviation from circularity.

To visualize ring orientation, we employ a rod representation. The center of each rod is positioned at the ring’s center of mass, and its orientation is aligned with the eigenvector corresponding to the largest eigenvalue ($\lambda_1^2$) of the gyration tensor and the length of the rod is proportional to $\lambda_1$. Further, to incorporate additional information about shape change, we scale the magnitude of the largest eigenvector (or the rod length) by the change in asphericity $da$ over the observation window.

\subsection{Defects}

To partition space among ring polymers, we employed a monomer-mediated grid-based Voronoi scheme. The simulation box was discretized into a uniform $1000 \times 1000$ grid with spacing $\Delta x \approx \sigma$, ensured for all densities considered ($\rho > 0.092\,\sigma^{-2}$), so that each grid point is unambiguously associated with a single nearest monomer. For each grid point, the nearest monomer was identified under periodic boundary conditions using the minimum image convention, and each grid point was assigned the molecule identifier of its nearest monomer. Two polymers were defined as Voronoi neighbors if at least one pair of adjacent grid points, scanned under four-connectivity, belonged to different polymers. This approach avoids center-of-mass generator points, which become unreliable when rings deform significantly under compression.

The coordination number $z$ of each polymer was defined as the number of distinct Voronoi neighbors. Topological defects were identified via the charge $q = 6 - z$, measuring deviation from the reference coordination of a hexagonal lattice~\cite{halperin1978theory}. A $5$--$7$ pair ($|q| = 1$) was counted as a single defect and a $4$--$8$ pair ($|q| = 2$) as a double defect. The total defect content was quantified as $\sum_{i}|q_i|/2$, summed over all polymers with $z \neq 6$.


\bibliography{poly}
\end{document}